\documentclass[preprint]{aastex6}

\usepackage{multirow}
\usepackage{hhline}
\usepackage{graphicx}
\usepackage{float}
\usepackage[caption = false]{subfig}

\begin{document}


\title{Observations of the Formation, Development, and Structure of a Current Sheet in an Eruptive Solar Flare}


\author{Daniel B. Seaton\altaffilmark{1,2},
  Allison E. Bartz\altaffilmark{3}, \& Jonathan M. Darnel\altaffilmark{1,2}}

\altaffiltext{1}{Cooperative Institute for Research in Environmental
  Sciences, University of Colorado at Boulder, Boulder, Colorado,
  80305, USA}
\altaffiltext{2}{National Centers for Environmental Information, National Oceanic and Atmospheric Administration, Boulder, Colorado, 80305, USA}
\altaffiltext{3}{Department of Physics, Grinnell College, Grinnell, Iowa, 50112, USA}
\email{daniel.seaton@noaa.gov}

\begin{abstract}
We present AIA observations of a structure we interpret as a current sheet
associated with an X4.9 flare and coronal mass ejection that occurred
on 2014~February~25 in NOAA Active Region 11990. We characterize the
properties of the current sheet, finding that the sheet remains on the
order of a few thousand km thick for much of the duration of the
event and that its temperature generally ranged between 8--10~MK. We
also note the presence of other phenomena believed to be associated
with magnetic reconnection in current sheets, including supra-arcade
downflows and shrinking loops. We estimate that the rate of
reconnection during the event was $M_{A} \approx$~0.004--0.007, a
value consistent with model predictions. We conclude with a discussion
of the implications of this event for reconnection-based eruption models.
\end{abstract}




\section{Introduction} \label{sec:intro}

Magnetic reconnection, the topological restructuring of magnetic field
lines embedded in a highly conducting plasma, is widely accepted to be
the mechanism that converts stored magnetic energy into heat and
kinetic energy during eruptive solar flares. In the standard model
\citep{1976SoPh...50...85K} a current sheet forms between antiparallel
field lines that are stretched outwards from the Sun in the wake of an
eruption. Reconnection in this current sheet releases magnetic energy
that powers a solar flare and simultaneously reconfigures the field to
form the post-flare loops and flare ribbons  commonly associated with
these types of eruptive events \citep{1989SoPh..120..285F}.

Reconnection-based flare models can explain a wide variety of
phenomena that are observed in association with eruptive flare current
sheets, including shrinking loops \citep{1996ApJ...459..330F,
  2008ApJ...675..868R} and inflows in the vicinity of flares \citep{1999ApJ...519L..93M,
  2009ApJ...697.1569M}. Reconnection models, meanwhile, have been
refined since \citeauthor{1976SoPh...50...85K}'s seminal paper, and
capture the dynamics and phenomena associated with these events with
ever-improving accuracy \citep[see, for
example,][]{2000JGR...105.2375L}. There are numerous reports of
  observations of reconnection-related phenomena in the literature, many of which are
  discussed below. However, there is still a need for direct
  observational confirmation of predictions from theory and numerical
  models, many of which remain poorly constrained. There is a
  particular need for observations that can help constrain properties
  of reconnection itself and, in turn, help us understand more
  precisely how reconnection liberates stored magnetic energy during
  eruptions. 

Of special importance are observations that can put limits on the
reconnection rate, which is a free parameter in many
models. The indeterminacy in reconnection models is discussed in
detail by \citet{2013PhPl...20e2902F}, who point out that the Petschek
model for reconnection, which itself is at the heart of many
reconnection-based eruption models, does not make a specific prediction
of the reconnection rate. So the need for observational constraints is
acute.

Likewise, other aspects of models can also be constrained by
observation. \citet{1997ApJ...474L..61Y}, \citet{2001ApJ...549.1160Y},
and \citet{2009ApJ...701..348S} all discuss the role of thermal
conduction in the energetics of reconnection in eruptive flares, but
there have been few observations that might allow us to assess how
much conduction actually influences the evolution of reconnection in
eruptive events. This in spite of the fact that new models are being
developed that take such effects into account
\citep{2013ApJ...776...54S}.

Cusp-shaped structures that form in the wake of solar flares are now
widely understood to be important evidence of reconnection as
well. This realization is present in the earliest discussions of such a model
\citep{1976SoPh...50...85K} and there are now numerous reports of such
structures in observations during solar flares and eruptions that can
be associated with both reconnection and
current sheets. (See \citealt{2015SoPh..290.2211G} for one such
example; for an extensive list of such observations, see Table~2 in
\citealt{2015SSRv..194..237L}.)

There are also reports of observations of flare-associated current
sheets and related phenomena using a variety of instruments and
wavelengths. These range from detections in hard x-rays
\citep{2013ApJ...777...93S} using the Reuven Ramaty High-Energy Solar
Spectroscopic Imager \citep[RHESSI;][]{2002SoPh..210....3L} to radio
wavelengths \citep{2013AandA...555A..40A}. In fact, the list of such
observations is far too long to discuss in any detail in this short
introduction; for a much more complete overview readers may wish to
consult \citet{2015SSRv..194..237L}, who discuss many other
observations of current sheets, the properties of the observed sheets,
and other associated phenomena. 

Here we focus on several specific observations that, like the one we
report in this paper, showed direct evidence of the current sheet
itself and provided the opportunity to test predictions by flare
reconnection models or to infer parameters that could be used to
constrain those models. 

Before we continue, it is worth pointing out that there is not a consensus on what the size of a current sheet in
  the solar corona is likely to be; predictions range from Larmor-radius scales
  of just a few meters to megameters. Nor for that matter, is there
  consensus on whether current sheets can actually be identified
  observationally and separated from other related, larger-scale
  structures\footnote{Good discussions of these diverse
    interpretations appear in the introduction of \citet{2013ApJ...766.39M} and 
 Section~4 of \citet{2015SSRv..194..237L}.} Different authors have dealt
  with this observational ambiguity in different ways, and
  correspondingly have referred to structures in their observations
  that might be related to current sheets with a variety of
  terminology. In this paper we will adopt a simple convention:
  we refer to the entirety of the spine-like structure in our extreme-ultraviolet
  (EUV) observations simply as the ``current sheet'', and our use of
  this term throughout the paper more generally refers to long, narrow structures
  on megameter scales that have been identified observationally by a
  variety of authors.

For example, \citet{2011ApJ...727L..52R} reported
observations from the {\it Atmospheric Imaging Assembly}
\citep[AIA;][]{2012SoPh..275...17L} on NASA's
{\it Solar Dynamics Observatory}
\citep[SDO;][]{2012SoPh..275....3P} of high temperature
reconnection-related plasma during an off-limb 
eruptive flare that occurred on 2010~November~3. Later analysis by
\citet{2011ApJ...732L..25C} and \citet{0004-637X-754-1-13} provided
further support to this hypothesis.

One of the clearest observations of a current sheet was reported by
\citet{2010ApJ...722..329S}, who described a very
narrow structure above a post-flare arcade system in observations from
the {\it X-Ray Telescope} \citep[XRT;][]{2007SoPh..243...63G} on-board
the {\it Hinode} spacecraft following the so-called “Cartwheel CME”
that occurred on 2008~April~9. These authors reported the width of this
structure to be 4--5$\times10^3\,\mathrm{km}$ and discussed its
association with other phenomena --- inflows and shrinking loops --- that
are believed to be signatures of magnetic reconnection associated with
eruptive flares.

A more recent paper by \citet{2016ApJ...821L..29Z} presented the
analysis of a current sheet that formed during a C2.0
flare. \citeauthor{2016ApJ...821L..29Z} observed a hot, extended
structure that formed at the top of cusp-shaped loops during the event
and showed, like \citeauthor{2010ApJ...722..329S}, that the structure was associated with
the type of inflows and outflows predicted by reconnection models. On
the other hand, unlike the current sheet described in the report by
\citeauthor{2010ApJ...722..329S}, this one was imaged in front of the
solar disk instead of in profile on the limb. This perspective afforded
\citeauthor{2016ApJ...821L..29Z} a good view of the conditions in the
region at the onset of the event, from which they concluded that this
current sheet --- although fundamentally similar to the fast growing
sheets observed during eruptive flares by other teams --- was likely
generated by a loop-loop interaction and may have grown considerably
more slowly as a result.  

There have also been reports of observations of suspected current
sheets in white light images that provided similar opportunities to
characterize important parameters. \citet{2014ApJ...784...91L}
described observations from the ground-based Mk4 K-Coronameter at the
Mauna Loa Solar Observatory \citep{2003SPIE.4843...66E} in the wake of
a CME on 2005~September~7. This eruption was associated with a flare
that was classified at least X17 (the flare saturated the X-Ray
Sensor on GOES, so its exact brightness is not known) and a CME
with speeds in excess of 2500 km/s. These authors measured properties of the
current sheet at considerably larger heights than the measurements
reported by \citet{2010ApJ...722..329S}, and found that the width of
the current sheet was roughly $2\times10^4\,\mathrm{km}$ out to
heights as large as nearly 3~solar-radii. 

In general, observations of current sheets such as these provide an opportunity to estimate the
rate of magnetic reconnection during the event, which is usually
reported in terms of the inflow Alfv\'en Mach number, $M_A$. $M_A$, in turn, is
proportional both to the ratio of inflow to outflow velocities,
$V_{in}/V_{out}$ and the ratio of width to length of the sheet, $l/L$
\citep[see][]{2000mare.book.....P}. \citet{2010ApJ...722..329S} used
the second ratio, based on the geometric properties of the sheet they
observed, to determine a reconnection rate between $0.002-0.006$,
while \citeauthor{2014ApJ...784...91L}, using the same method, found
values of $M_A$ closer to 0.07, a number more consistent with
predictions from many models. \citet{2000JGR...105.2375L}, for
example, describe a model with  $M_A = 0.1$, though it is worth
pointing out that they note a rate as low as $M_A = 0.005$ can still
yield an eruption. \citeauthor{2016ApJ...821L..29Z} measured the reconnection rate
by estimating flow velocities and found $M_A$ between about 0.01 and
0.05. 

The reconnection rate is an especially important property for constraining
reconnection-based eruption models because it determines the rate at
which magnetic energy can be converted to heat and kinetic energy to
drive a solar eruption and flare. Exactly what determines rate of
reconnection in the corona and, as we mentioned above, what the rate
  is predicted to be, remains the subject of much research and
  discussion \citep{2013PhPl...20e2902F}.

It is also worth noting that current sheets in the corona are complex,
three-dimensional structures whose appearances are determined by many
factors, especially the perspective from which they are
observed. Accordingly, there have been many other reports of
current-sheet-associated phenomena that do not fit the classical,
two-dimensional cross-sectional picture. When viewed face-on, current
sheets are often described as ``fans'' \citep[see, for example,][]{2007AandA...475..333K,
  2011ApJ...730...98S, 2011ApJ...742...92W}. From this perspective,
the current sheet is a broad, turbulent, high-temperature curtain
punctuated by fast inflowing loops that often trail dark voids, commonly referred to as supra-arcade downflows (SADs) in their wake
\citep{2012ApJ...747L..40S}. Observations of fans and inflows have been made
using a variety of imagers --- for example, the {\it Yohkoh Soft X-Ray Telescope}
\citep{1999ApJ...519L..93M}, the {\it Transition Region and Coronal
  Explorer} \citep{2002SoPh..210..341G}, XRT \citet{2015SSRv..194..237L}, and AIA
\citep{2012ApJ...747L..40S} --- and have been believed to be linked to reconnection
since their first detection during the Yohkoh mission nearly two
decades ago. As we will discuss later in this paper, these
manifestations are part and parcel of the same current sheet phenomena
observed in this and other events. One peculiarity of the event
we discuss in this paper is that the three-dimensional nature of the
sheet provides the opportunity to study both edge-on and face-on
phenomena simultaneously. 

Observations of face-on phenomena like SADs, too, can provide an
opportunity to evaluate model
predictions. \citet{2014ApJ...786...95H}, for example, used
differential emission measure (DEM) analysis to characterize the
temperature structure of a number of SADs and the surrounding plasma,
and found that the current sheet plasma in the fans often had
temperatures in excess of $10^7$~K, and SADs, while hot, were
generally cooler than the surrounding fan.

Likewise, in an analysis of similar events \citet{2016ApJ...819...56S}
attempted to deduce the magnetic structure of SADs and the plasma that
surrounds them. In particular, they attempted to determine the plasma beta, which is
the ratio between thermal and magnetic energy density in a
plasma. Beta is an important parameter in reconnection models, and is
often assumed to be very small, as it is in most of the low
corona. Their analysis showed that this assumption is likely
incorrect, providing another important constraint for theory and
models.

In this paper we present AIA observations of an X4.9 flare that occurred
on 2014~February~25 that contain evidence of the presence of an
extended, high-temperature current sheet that formed in the wake of an
eruption and was clearly associated with the growth of the post-flare
loop system. We attempt to measure or detect several properties
  of the sheet structure that might be useful in constraining
  reconnection and eruption models, including the reconnection rate
  and the presence of a thermal halo that might be linked to
  conduction from the sheet into the surrounding corona. We also
discuss its association with commonly observed eruptive flare-related
phenomena like supra-arcade downflows and shrinking loops. We conclude
with a few remarks about what can be learned from observations like these
and how research into the nature of magnetic reconnection during solar
eruptions might be enhanced by future instrumentation.

\section{Observations} \label{sec:obs}

\begin{figure}
\centering
\includegraphics{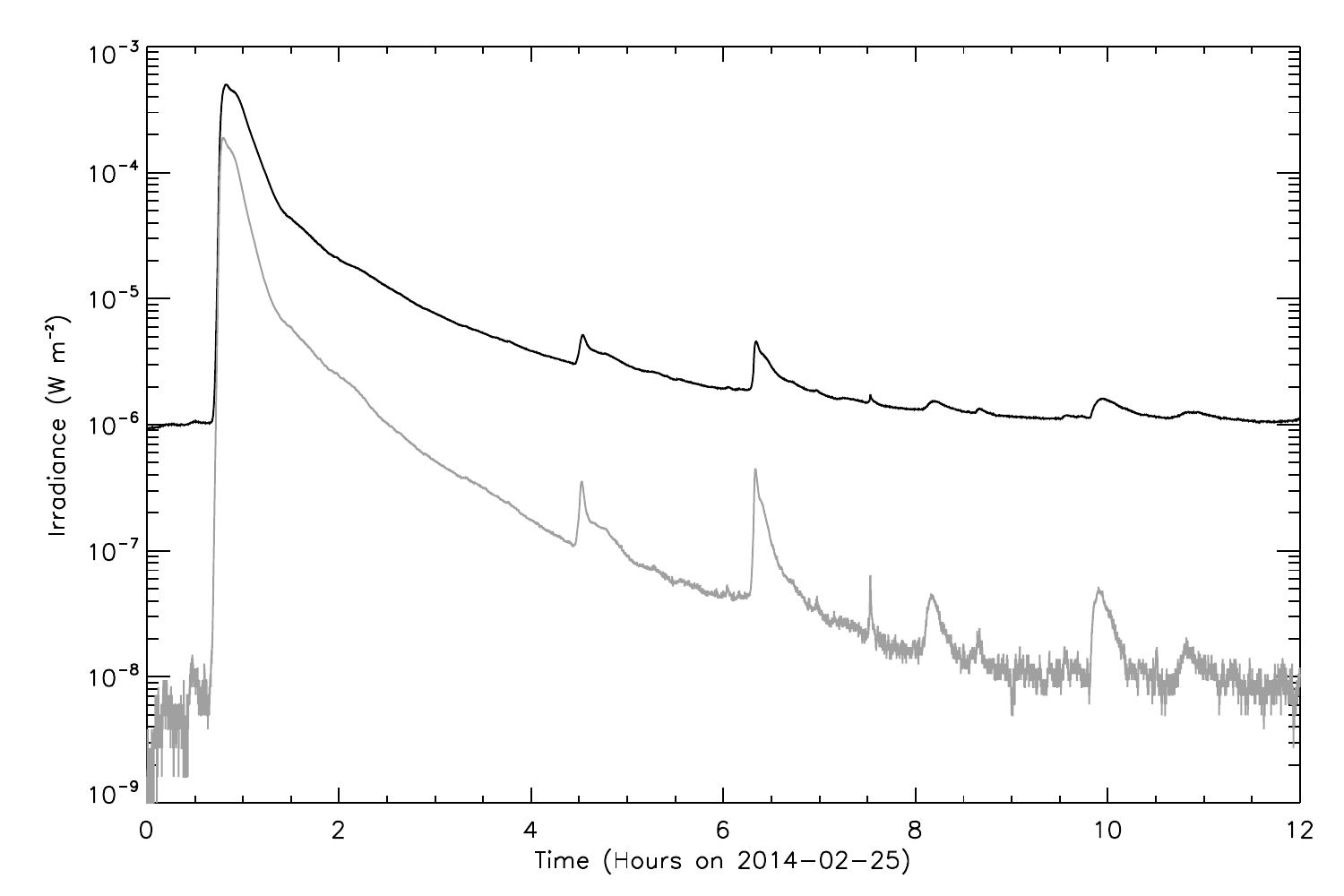}
\caption{X-ray irradiance measured by the GOES X-Ray Sensor
  instrument, beginning at midnight on 2014~February~25. The black
  curve corresponds to the 1--8~\AA\ channel while the gray curve shows the
  0.5--4~\AA\ channel. }
\label{fig:goes-flux}
\end{figure}

The structures we report on in this paper were formed in association
with a large and complex filament eruption that occurred on the east
limb of the Sun in NOAA Active Region 11990 at about 00:40~UT on
2014~February~25. This eruption was also associated with an impulsive
X4.9 class flare, which peaked about 10~minutes after the onset of the
eruption, and a very energetic CME with a reported velocity of more
than 2100~km/s in the CDAW CME Catalog \citep[for a description of
CDAW, see][]{2004JGRA..109.7105Y}. X-ray irradiance measurements of the  
flare from the GOES spacecraft show a nearly instantaneous rise to the
peak brightness from the onset of the event, and a long and gradual
decay to background levels some 6--8~hours after the event's
onset. Figure~\ref{fig:goes-flux} shows a plot of the measured
irradiance during the event.

A previous analysis of this event by \citet{2014ApJ...797L..15C} presented
evidence that the filament eruption itself was triggered by tether-cutting
reconnection at the base of the erupting filament. These authors did
not, however, examine the role that reconnection played in the
development of structures that formed later in the event's evolution. 

In this paper we focus on a structure, which we interpret as the
manifestation of a current sheet, that appeared only after the
eruption's onset and continued to evolve for at least six hours,
throughout the decay phase of the flare. We studied this event using
observations from the AIA and, although we used observations from all of AIA's available
extreme-ultraviolet channels, we focused in particular on observations
in the 131~\AA\ channel because its response to high-temperature
Fe~\textsc{XXI} \citep{2010AandA...521A..21O} provided the best view of
the current sheet structure that is the subject of this paper. 

The feature of interest in this paper formed in the wake of the
erupting filament that was visible in AIA images between 00:40 and
00:52. The structure first appears as a very narrow spine, growing
outward from the flare site between 00:59 and 01:04. This structure
remains very thin for nearly an hour, until about 01:40, when it
begins to widen. By 1:47, a fainter cusp-shaped structure appears at
the base of the spine, above the post-eruptive loops associated with
the flare. The cusp broadens over time until it reaches its maximum width
at around 03:40. Finally, the cusp structure narrows and becomes fainter over
time, but remains distinct from the background until about 08:00, while
the spine structure disappears progressively downwards into the cusp structure by
about 05:50.

\begin{figure}
\centering
\includegraphics{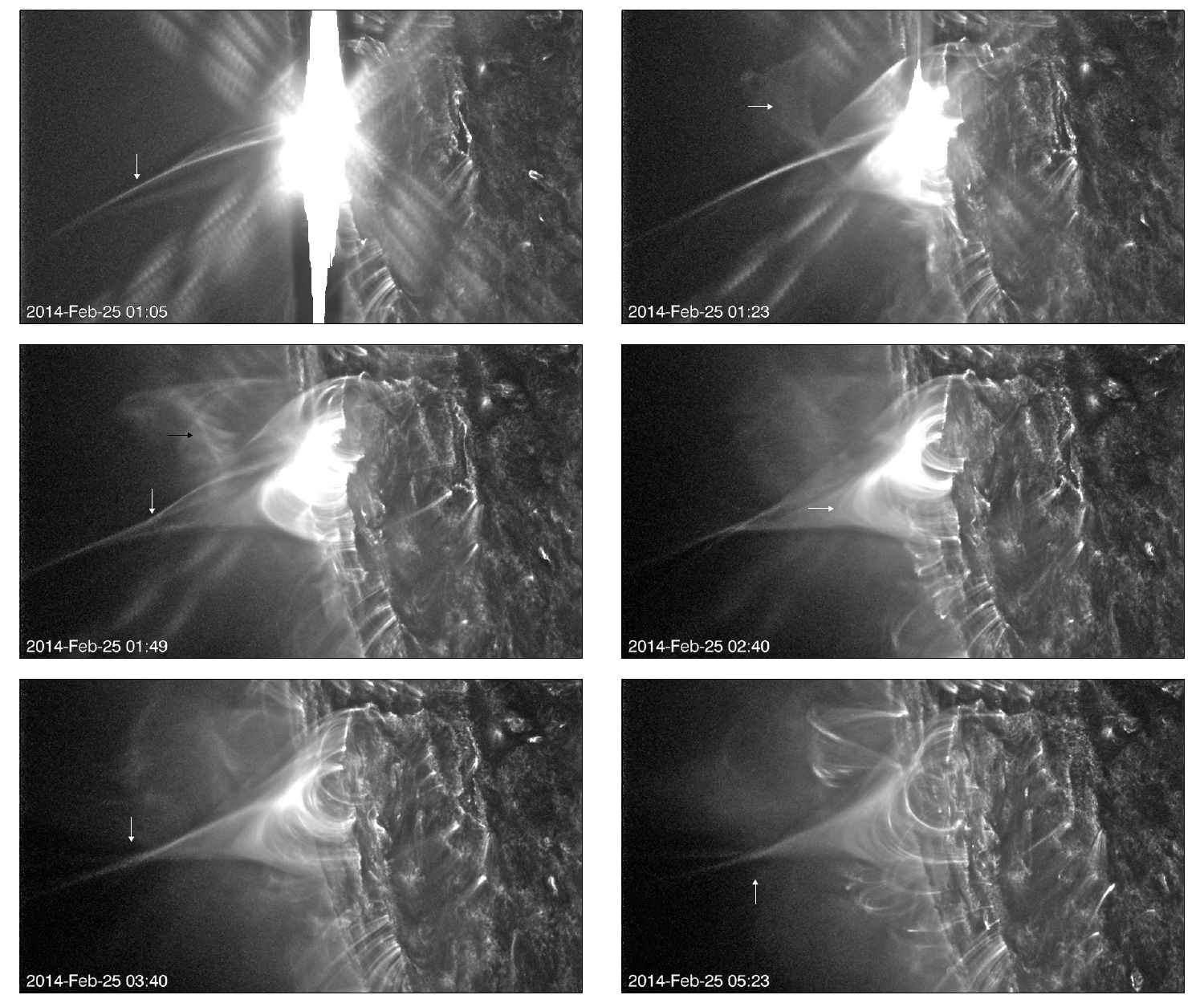}
\caption{The evolution of current sheet structure in the 131~\AA\ AIA
  channel beginning with its appearance in the wake of a strong
  CME. Early on (upper left) the structure is long and narrow, and
  only later (upper right) do background features begin to
  appear. These features are first seen as shrinking loops, which
  later broaden (middle left, black arrow) into a more fan-like
  structure while the sheet itself (middle left, white arrow) begins
  to broaden. As the sheet broadens, shrinking loops are clearly
  visible in the cusp region at the current sheet's base (middle
  right). Even later (lower left) dark inflows, presumably SADs,
  become visible in the diffuse background emission. At very late
  times (lower right) some material appears to flow into the sheet
  itself, triggering bifurcated up-down flows along the sheet
  structure. An animation of the evolution of this structure is
  included with the online version of this paper.}
\label{fig:evolution}
\end{figure}

Figure~\ref{fig:evolution} and the accompanying animation show the
evolution of the flare region over the course of the
event. To generate this sequence of images we first eliminated underexposed
images generated by AIA's automatic flare exposure control, which
produces short-exposure images during bright flares. Although this
exposure control helps to limit the amount of saturation and blooming
due to the flare itself, the resulting images did not provide a good
view of the current sheet structure. 

Many of the dynamic features of this event are the result of very weak
variations in already faint signals, so we further processed these
images to improve the visibility of these features and reduce
noise. We began by median-stacking them in groups of seven, which appeared to
yield the best balance between avoiding motion blur while controlling
image noise. We then applied a routine included in the {\sf SolarSoft
  IDL} \citep{1998SoPh..182..497F} branch for the SWAP instrument
\citep{2013SoPh..286...43S} on-board PROBA2 called {\sf
  gentle\_filter.pro}. This algorithm is a noise filter that
suppresses individual pixels that are strong outliers compared to their neighbors without
significantly reducing image sharpness; we applied it to the resulting images to
further remove speckle noise in the images. Finally, we applied an
image filter that improved local image contrast by normalizing the
image brightness using a function derived from a gaussian-smoothed copy of the
image. The resulting images were combined into a movie in which
large-scale features are preserved while the visibility of local
variations is enhanced.

The six different panels of Figure~\ref{fig:evolution} highlight several important
phenomena that we observed during the event. The first panel (top left) at
01:05, shows the early formation of the current sheet, which is
highlighted by the white arrow. By 01:23 (top right) the current sheet structure
is well-formed and there are indications of a coronal fan structure
forming as well, indicated again by a white arrow. This fan is almost
certainly related to the current sheet, but is seen face-on rather
than edge-on thanks to the complicated three-dimensional nature of
this event.

By 01:49, the time of the third panel (middle left) the bottom of the sheet has
begun to broaden as described above (white arrow), while the fan has
become much more visible and highly structured (black arrow). By 02:40
(middle right) we see many loops shrinking through the broad,
cusp-shaped feature at the bottom of the sheet, a clear indication
that flare-related reconnection is probably taking place. 

Throughout this time we also see evidence of SADs in the fan both to
the north and south of the sheet. The white arrow in the panel at
03:40 (bottom left) indicates one example of such an
inflow. These inflows are another common phenomenon associated with
flare reconnection, and are closely linked with the flows we see along
the sheet itself. This is yet another indication that the fan and
sheet are part of the same three-dimensional structure viewed at
different angles.

By the time of the final panel (bottom right) at 05:23 there are
indications not only of downflows from above, but also of flows into
the current sheet itself. The white arrow highlights one such example,
which appears to be advected into the sheet before bifurcating into
upward- and downward-directed flows. This may be an indication of
ongoing magnetic reconnection, or even of the presence of an magnetic
neutral point or flow stagnation point in the sheet. Unfortunately,
due to the weak contrast of these features, we could not
make quantitative measurements of the nature of these flows. Because
of this, and because of the highly turbulent nature of this region more
generally, we hesitate to draw any strong conclusions about the
fundamental nature of these flows.

\section{Analysis} \label{sec:analysis}

\subsection{The structure of the current sheet} \label{subsec:struct}

It is worth emphasizing once more the complex, three-dimensional nature of the
structures associated with this eruption. From the perspective of SDO,
we see post-eruptive loops both from the edge-on perspective, as in
those highlighted in the panel of Figure~\ref{fig:evolution} at
01:49~UT, and in a face-on perspective, like the shrinking loops that
appear in the panel at 02:40. Here we focus our analysis on the
spine-like current sheet structure that overlies
the shrinking loops, because it provides a rare opportunity to deduce
some of the properties of eruption-associated magnetic reconnection
directly from observations.

However, the association of this feature with the more frequently
observed SADs and coronal fan structure is also important. Although
SADs and fans have long been understood to be the manifestation of
current sheets viewed from the face-on perspective, there have been
very few observations that clearly reveal both aspects of this
phenomenon simultaneously. This event provides a unique opportunity
for direct confirmation of the relationship between these phenomena.

The focus of our analysis here was to determine properties of
the spine-like current sheet structure that might help us characterize
the processes that underlie both the structure's formation and the
eruption itself. We measured the width of the sheetlike structure
using a combination of algorithmic image processing and visual
inspection. We first used an automated algorithm to determine the
brightest point in every column the feature spanned. We then fit the
resulting points using a piecewise-defined linear regression fit such that
the changing slope of the feature over its length was well-captured by the fit.

We used the slope from the linear fit to define an intersecting sample
of 100 pixels perpendicular to the feature. Because the thickness of
the structure varied somewhat, we selected the location of the point of
measurement with respect to each subframe to obtain a representative
value for the thickness.

We sampled the background brightness by computing the mean of a small
region of pixels near --- but not including --- the structure.  We then
iteratively rejected pixels which were not brighter than the
background value times a constant, progressively increasing the
constant until no background points remained that were clearly beyond
the edge of the feature, as determined by visual inspection. We then computed the
width of the feature at various points along its length by measuring
the distance between  between the first and last of the remaining
points. We consider this measurement to be the upper limit on the
width of the feature.

\begin{figure}
  \centering
  \subfloat{\includegraphics[height=2.75in]{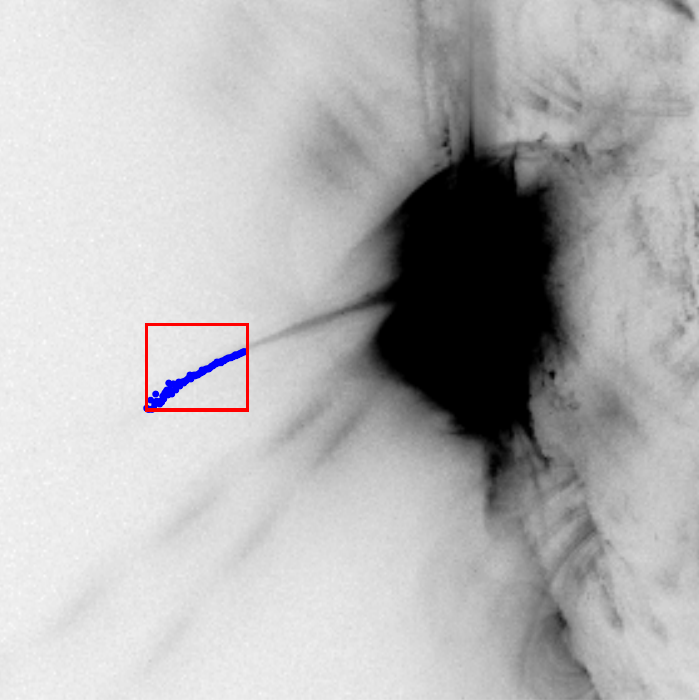}\hspace{0.125in}}
  \subfloat{\hspace{0.125in}\includegraphics[height=2.75in]{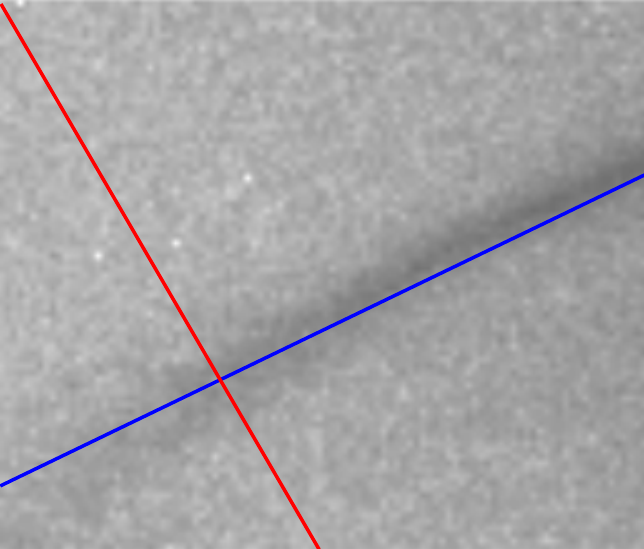}}\\
  \subfloat{\includegraphics[width=6.8in,height=3.75in]{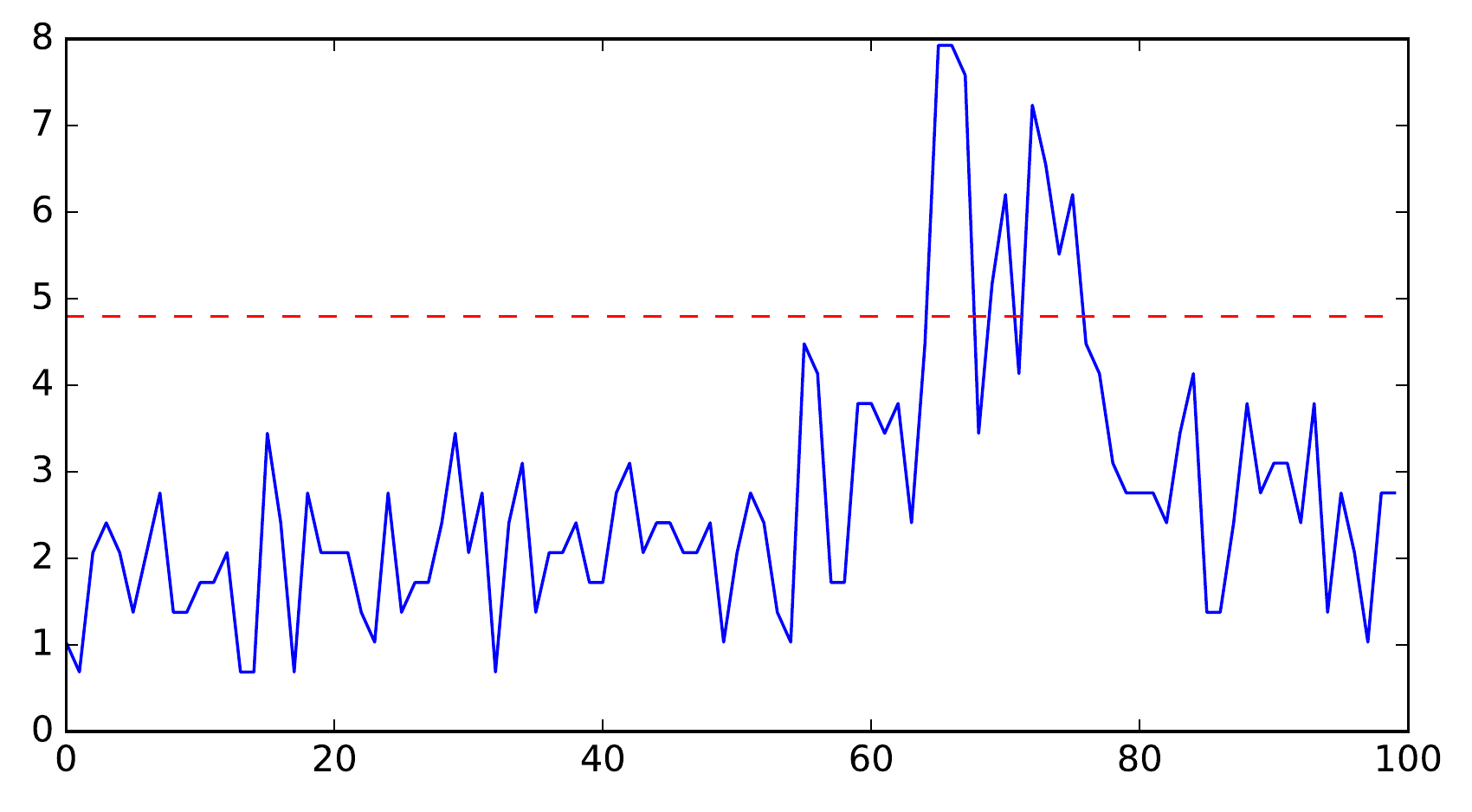}} 
\caption{An illustrative overview of the method we used to measure current sheet
  widths. We first identified the location of the sheet in a
  sub-region by locating the brightest pixel in each column (upper
  left). We then fit these points with a line and sampled a
  cross-section of the current sheet feature perpendicular to this
  line (upper right). We identified the width by determining a
  background noise floor value above which the outermost edges of the
  feature were clearly part of the current sheet (bottom).}
\label{fig:width_algor}
\end{figure}

In many cases, however, the width of the feature is ambiguous, since
its edges are poorly defined.  We thus determined a lower limit on the
width of the feature by increasing the background value
further until the remaining points barely spanned the feature, as
determined by visual inspection. Figure~\ref{fig:width_algor} shows an
illustration of the approach we used to measure widths of the current
sheet at various heights, the results are reported numerically in
Table~\ref{table:width}. The values of width reported in the table are the mean
of the upper and lower limit for the various points at which we
measured the feature, while the reported error is the mean of the
difference between these limits and the reported width.

We generally considered the maximum width to be the widest
possible measurement where no pixel clearly extended into the
background, and the minimum to be the thinnest measurement that
completely spanned the feature. For example, the base of the feature in the
image from 03:20:44 has a cusp-like structure with the more or less uniform
brightness, the width of which we considered to be the minimum, as
well as a fainter extension of the cusp extending along the north side
of the feature, which was included in the maximum. 

\begin{figure}
\centering
\includegraphics[width=1\textwidth,natwidth=23.75]{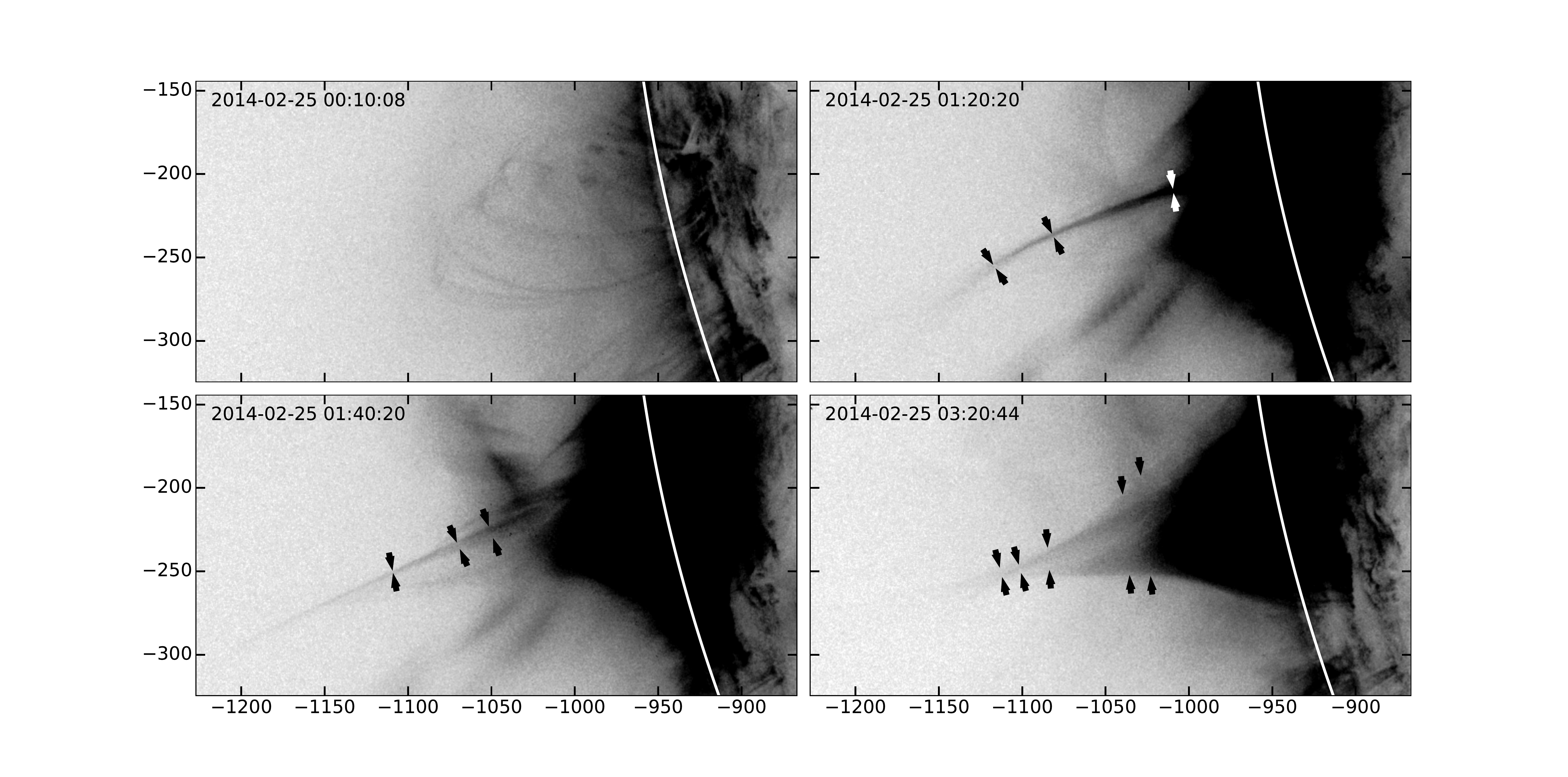}
\caption{Measured widths of the current sheet at various times and
  locations during its evolution. The panel in the upper left, at
  00:10:08~UT, from before the eruption onset, is included to show
  that there is relatively little pre-existing background structure that might
  influence our ability to accurately measure the sheet width later on.}
\label{fig:width}
\end{figure}

We performed this measurement on images obtained at 1:20:20, 1:40:20,
and 3:20:44. Figure~\ref{fig:width} shows where width measurements
were taken (indicated by arrows) throughout the evolution of the
structure. The image from 00:10:08 shows the region before the
eruption, and confirms there are few strong background features that
might interfere with measurements in the later images. Although
all of the images in the figure are displayed using a power-law scaling
with a gamma of 0.4, the data used in our width measurements were not
scaled. It is worth noting that the images at 1:40:20 and 3:20:44 were created by
computing the median-value of pixels from a stack of three images: an
image at the given time, an image immediately preceding the given
time, and an image immediately following the given time, to help
reduce detector-related noise.

\begin{table}
\begin{center}
\caption{Measured current sheet width and error at various heights and
  times in the evolution of the event.} 
\begin{tabular}{  r | r | r | r | } 
\multicolumn{1}{r}{}
   &  \multicolumn{1}{c}{Height (solar-radii)}
   & \multicolumn{1}{c}{Width (km)} 
   & \multicolumn{1}{c}{Error (km)} \\
\hhline{~---}
01:20:20 & 1.181 & 3880 & 860 \\\hhline{~---}
  & 1.142 & 1720 & 430 \\\hhline{~---}
  & 1.064 & 2800 & 220 \\\hhline{~===}
01:40:20 & 1.173 & 1720 & 430 \\\hhline{~---}
  & 1.129 & 3280 & 950 \\\hhline{~---}
  & 1.108 & 4840 & 1560 \\\hhline{~===}
03:20:44 & 1.177 & 3980 & 1360 \\\hhline{~---}
  & 1.165 & 4710 & 400 \\\hhline{~---}
  & 1.146 & 10200 & 640 \\\hhline{~---}
  & 1.098 & 31800 & 4100 \\\hhline{~---}
  & 1.083 & 38300 & 6290 \\\hhline{~---}

\end{tabular}
\end{center}
\label{table:width}
\end{table}

At this point is is worthwhile to ask whether these measurements
  are really likely to provide an accurate assessment of the width of
  the current sheet. As we will see later in this paper, this is an important
  question, since the width of the sheet is an essential input to the
  calculation of the reconnection rate. Likewise, the width of this
  structure --- and any extent it has beyond the current sheet itself
  --- also has implications for our ability to use these observations
  to probe for evidence that might help clarify the role of thermal
  conduction in current sheets more generally.

Why might we over-estimate the width of the current sheet? One straightforward
possibility is instrumental error. AIA has an effective resolution of approximately 1.7~arcsec (or about
three pixels) in the 131~\AA\ channel \citep{2012SoPh..275...41B}. The structures we
describe here are tens of pixels wide and the error resulting from
our measurement technique alone is, in general, several
pixels. Thus instrumental effects are not likely a cause of
such an over-estimate any more than our ability to accurately locate the edge
of the bright structure in the observations.

This means that any over-estimate would have its roots in the physical
properties of the structure, rather than in measurement error. It is worth
noting that our estimates match those of many others, including several
of those discussed above in Section~\ref{sec:intro}. It is also worth
pointing out that \citet{2015SSRv..194..237L} examined this question
in detail and concluded observational estimates of widths of current
sheets like the one we present here probably do not differ substantially from
the true physical widths of these structures in the first place.

We might also wonder why, since flare reconnection is widely understood to be a patchy,
turbulent process \citep[see][for one discussion]{2012SSRv..173..557L}
our current sheet can be observed as a persistently laminar structure
for long enough for us to accurately measure its width at all. In fact, the
sheet is not laminar in general. As we discuss in
Section~\ref{sec:obs}, its later evolution is characterized by complex
inflows and outflows, clear evidence of shrinking loops, and a much
less uniform appearance. 

Turbulence in current sheets is closely
linked to the onset of instabilities --- most importantly, the tearing
mode instability --- which can lead to the formation of magnetic
islands and the breakdown in laminarity of the reconnection outflow
from the tips of the sheet. Theoretical and numerical analyses differ
about what conditions under which onset of the tearing mode is likely to occur. However,
predictions agree that it is not likely to occur before the aspect
ratio of the sheet, its length divided by its width, reaches some
critical value. Recent analytical and numerical experiments by
\citet{2010PhPl...17e2109N} suggest that the critical value might be
somewhere between about 50 and 200. Since the aspect ratio of our
sheet during its early evolution (at least for the the first two times
reported in Table~\ref{table:width}) is around 200 (see
Section~\ref{subsec:reconrate}) we can reasonably conclude that the
sheet may appear stable because the instabilities that lead to
turbulence have not yet set in at the time of these measurements.

\subsection{Differential emission measure analysis} \label{subsec:DEM}

One question we hope to address with these observations is
  how well predictions using models that include thermal
  conduction capture the physics of a real current sheet in the
  corona. As we have noted previously, many authors have explored this question and
predicted the formation of a so-called ``thermal halo''. That is, they
predict that current sheets will be embedded in a much broader region
of warm plasma created by heat flow from inside the sheet into the
surrounding corona.

To address this question we attempted to characterize the temperature
profile of the structure and its surroundings using DEM analysis. To do this, we
used a version of the {\sf SolarSoft} code {\sf xrt\_dem\_iterative2.pro}, which was
designed for DEM analysis using observations from the X-Ray Telescope
(XRT) on Hinode \citep{2004ASPC..325..217G, 2004IAUS..223..321W,
2012ApJ...761...62C}, modified to accept observations from AIA instead. 

To ensure that our DEM results were not contaminated by background
illumination from the very bright core of the flare being spread
by AIA's instrumental point-spread-function (PSF), we first used the
{\sf SolarSoft} routine {\sf aia\_deconvolve\_richardsonlucy.pro} to
perform a 25 iteration deconvolution and
remove as much PSF-derived signal our region of interest as possible. 

Although the structure we hoped to characterize was quite obvious in
observations of the AIA 131~\AA\ and 94~\AA\ channels, it was still a
relatively faint structure superimposed on an even fainter
background. As a result, the region we hoped to study was strongly
affected by detector noise in many of the images we planned to analyze. To
improve the image statistics we took several steps. First, we co-added
frames taken in the same bandpass using pixel-by-pixel
  median-stacking on the exposure-time-normalized images to help
reduce spatial variations related to detector shot noise. For each channel we used images
obtained at 12-second cadence over approximately 2.5-minute windows,
which yielded the maximum number of images we could stack without
introducing significant artifacts due to motion or other changes
during the intervening period. 

Note that, because of the AIA automatic exposure control in the
  wake of the flare, some of the images we used were under-exposed ---
  particularly at large heights above the limb --- 
  while others contained a saturated region near the flare
  core. Because our median-stacking approach excludes outlying values,
  it splits the difference, and our stacked frames are generally free
  of both saturated and undervalued pixels. This yields frames with
  the best possible signal-to-noise given the limitations of the input
  data. 

To improve signal-to-noise further, we $2\times2$~rebinned the
resulting stacked data, resulting in an overall reduction of
resolution but an improvement in counting statistics by a factor of
four in each pixel. 

Nonetheless, in the region surrounding the current sheet there was
still very little signal in several AIA channels, which meant our DEM
solutions were still relatively poorly converged in the vicinity of
the current sheet. For points in the region surrounding the current
sheet, peak temperatures were between 2--3~MK. Other authors who
  measured temperatures in this range in the vicinity of current-sheet
related phenomena \citep{2014ApJ...786...95H, 2016ApJ...819...56S}
attributed them to foreground and background emission along the line
of sight. 

In our case, this region also represents the three-dimensional extension of
the current sheet along the limb to the north and south of the
spine-like feature that we interpret as the current sheet seen
edge-on, so we wonder why the solutions in this region are not closer
to the temperatures measured in the edge-on current sheet itself. In this
case, the poor convergence of our DEM solutions in this region appears
to offer a clue: it is likely there is simply not enough emission measure
related to the fan structure --- and thus not enough signal --- to assess
the temperature of this region with great accuracy. 

\begin{figure}
\centering
\includegraphics[width=\textwidth]{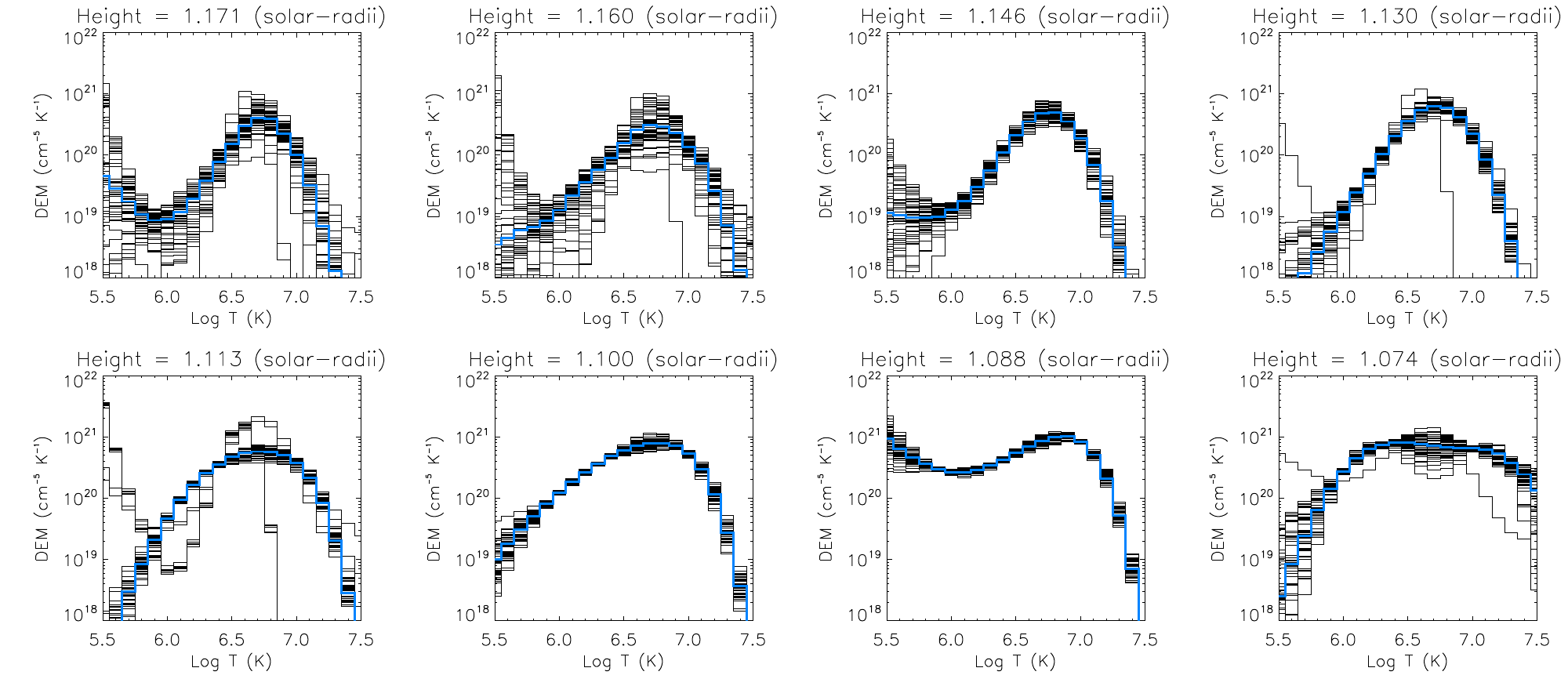}
\caption{DEM fits for a variety of heights along the current sheet at approximately
  01:28~UT. The blue curve indicates the best fit solution, while
  the black curves correspond to Monte Carlo iterated solutions that
  provide a general estimate of the inherent error in the DEM
  estimates plotted in blue. In general, fits higher in the corona
  (top panels) where the overall signal is weaker were noisier than
  those low in the corona.}
\label{fig:DEM}
\end{figure}

Our DEM solutions in the current sheet, where the signal-to-noise
ratio was stronger, were considerably better converged than solutions
outside this region. Figure~\ref{fig:DEM} shows reconstructed DEMs for
several representative points at various heights along the current layer, derived from
a stack of images obtained between 01:27:00 and 01:29:30. These solutions
all contained clear peaks at temperatures between 8--10~MK, an
unambiguous indication that the current sheet we studied here was a
largely uniform, high-temperature structure.

As the figure shows, in general the solutions are better converged
lower in the corona where the signal is stronger. DEM reconstructions
also show temperatures that are slightly higher --- by 1--2~MK --- at
lower heights, although the sheet is strikingly uniform over its full
length.

There are good reasons to be self-critical about the use of PSF
  deconvolution in our analysis, most importantly that this step might
  inadvertently alter the ratios of brightness in the current sheet
  region in deconvolved images in a way that alters the solutions. To
  assess this risk we performed all of the analysis steps described
  above using both deconvolved and unprocessed images and compared the
  results. We found that the solutions for the deconvolved images
  were, in general, better converged. The curves in the deconvolved
  solutions also tended to be weighted slightly more towards higher
  temperatures, but the difference is much smaller than the error
  inherent in the analysis itself, and in many cases the curves for
  both deconvolved and unprocessed images had the same peak
  temperature. So it appears the deconvolution did not strongly affect
  our analysis of the temperature of the current sheet.

The deconvolution had a much more significant effect in the background
region. By far, the strongest sources of PSF-derived signal are the
bright core of the flare near the base of the current sheet. Because
of the shape of AIA's PSF, signal originating from the flare
disproportionately leads to an increase in brightness in the
background region. Deconvolution removes the PSF-derived background
brightness that could interfere with our ability to identify a region
of enhanced temperature --- the so-called thermal halo --- surrounding
the current sheet. Since deconvolution does not appear to strongly
bias our results, we believe this step is worth taking.

A careful exploration of the DEMs in the current sheet and the
surrounding region reveal a highly uniform, hot current sheet and a
cooler background with very little evidence of a transition layer
between the two regions. This suggests that, if a thermal halo is
present, it is either too faint or too narrow a region to be
detected.

Predictions by \citet{1997ApJ...474L..61Y,
  2001ApJ...549.1160Y} suggest such a halo would be cooler than the
current layer by approximately a factor of two and perhaps as wide as
the sheet is thick. On the other hand, these predictions also suggest
a sheet with strong thermal conduction would itself be highly uniform
in temperature, as we do see in these observations, so the results
here are, unfortunately, a bit ambiguous. We will consider the
implications of this apparent mismatch with predictions in Section~\ref{sec:disc}.

\subsection{Reconnection rate} \label{subsec:reconrate}

In the simplest case --- two dimensional reconnection --- the
reconnection rate is simply the rate at which field lines move through
the x-type neutral point in the current sheet. For a simple,
two-dimensional configuration, this rate is given by $\partial A
/ \partial t$, where $A$ is the magnetic flux function, so according
to Faraday's equation, the reconnection rate is given by the strength
of the electric field at the neutral point where reconnection is
actually taking place \citep{2000mare.book.....P}.

In the case of a simple X-line, where the current density outside the
diffusion region where reconnection is taking place is very small, we
can approximate the electric field, $E_{0}$, by $E_{0} = M_A v_{Ae} B_e$, where
$M_A$ is the external Alfv\'en Mach number, $v_{Ae}$ is the 
external Alfv\'en speed, and $B_e$ is the external magnetic field
strength. Since in many models the field strength and Alfv\'en velocity
are taken to be fixed quantities and are often normalized out of the
system, the reconnection rate is usually parameterized by the
dimensionless value $M_A$.

As we discussed in the introduction, measuring $M_A$ is not
straightforward, but \citet{2000JASTP..62.1499F} argue that a good estimate
--- or at least, a good limit --- can be obtained from the ratio of sheet
thickness to length, which several authors have used to estimate
reconnection rates from observations of current
sheets.

Here we are able to very accurately measure the thickness of the
sheet, but obtaining an estimate of its length is not so
straightforward. Within a few minutes of its formation, the sheet
becomes long enough that its upper tip extends beyond the edge of
AIA's field of view.

Instead we turn to observations of the out-going eruption from the
Large Angle and Spectrometric Coronagraph \citep[LASCO;][]{1995SoPh..162..357B}
onboard the Solar and Heliospheric Observatory (SOHO)
spacecraft. Unfortunately, data from LASCO's inner C2 coronagraph
contains an observational gap just at the time of the onset of the
CME, so only a few frames contain useful observations from which to
estimate the length of the current sheet. One frame in particular,
from 01:25:50~UT, does appear to reveal the CME relatively early in
its evolution.

\begin{figure}
\centering
\includegraphics[width=\textwidth]{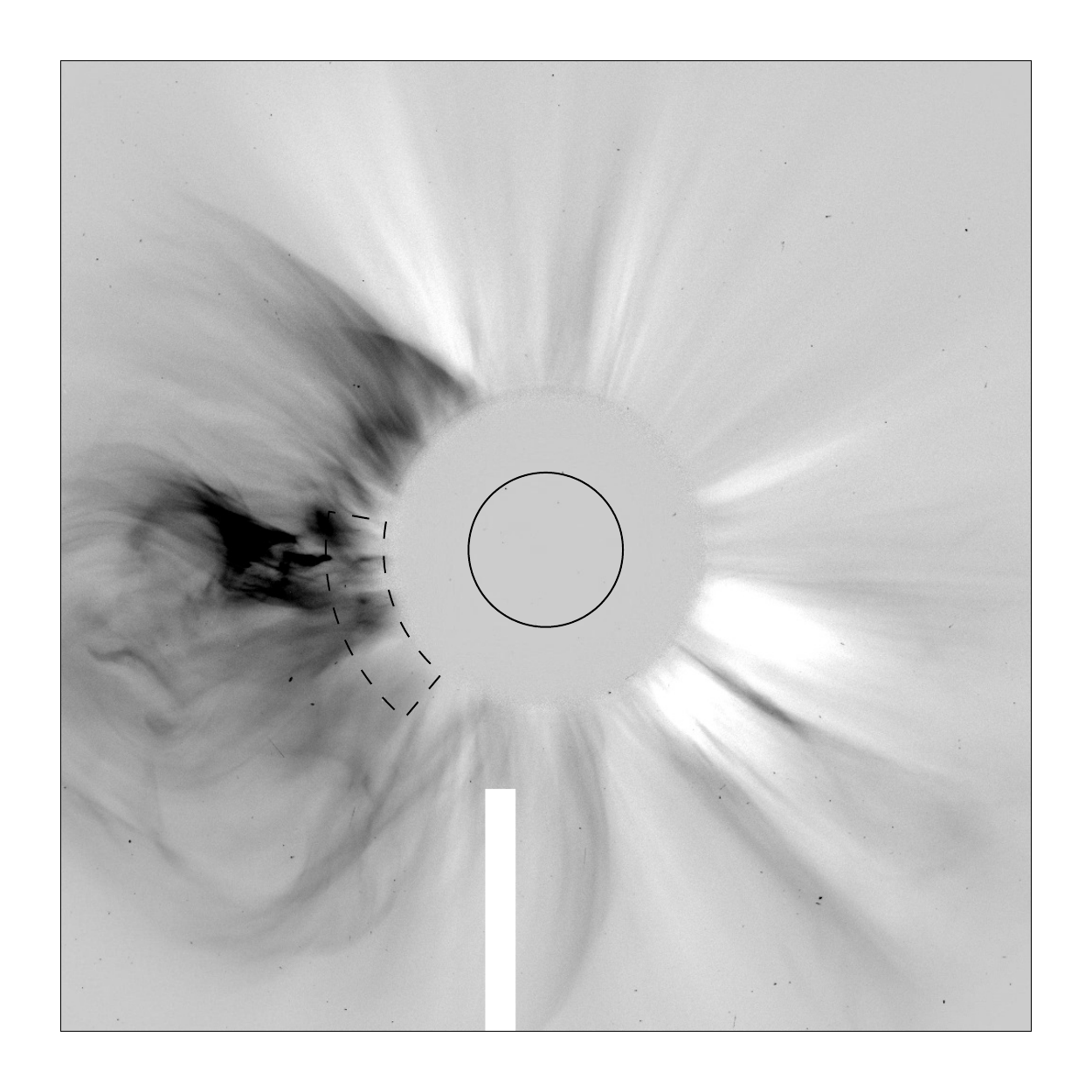}
\caption{A background-subtracted LASCO coronagraph image (inverse
  grayscale) obtained at 01:25:50~UT allows us to estimate the
  location of the upper tip of the current sheet (dashed box)
  associated with this eruption.}
\label{fig:LASCO}
\end{figure}

Figure~\ref{fig:LASCO} shows an image generated by subtracting the
median background brightness over the course of several hours before
and after the eruption from the frame obtained at 01:25. Circular structures associated
with the erupting flux rope are clearly visible to the east, while
closer to the occulting disk, inside the overlaid dashed box, we see
several structures that are suggestive of the sort of inverted,
cusp-like features predicted by many models \citep[like the one described in][for
example]{2000JGR...105.2375L} to mark the upper tip of current
sheets. There is likely no way to prove these structures are, in fact,
associated with the upper tip of the current sheet we observed using
AIA, but since the current sheet certainly cannot extend into the erupting flux
rope, this region provides a reasonable estimate for the location of
the upper tip nonetheless. 

The dashed box spans heights between 2.1 and 2.85~solar-radii. Since
the location of the lower tip of the current sheet is visible in AIA
images and is located between about 1.05 and~1.1 solar-radii at this
time, we can conclude the current sheet was between --- and certainly no
longer than --- about $7\times10^5$ and $1.2\times10^6$~km in length.

Taking the ratio of the estimated widths of a few thousand km
discussed in Section~\ref{subsec:struct} and these estimated lengths gives
a reconnection rate of $M_A \approx 0.004-0.007$. Although this rate
is relatively slow, it is similar to the rates reported by
\citet{2010ApJ...722..329S} and is reasonably close to
\citeauthor{2000JGR...105.2375L}'s estimate  that a minimum
reconnection rate of $M_A = 0.005$ would be required to power a strong
eruption and flare.

We emphasize that these values provide
only an estimate of the real reconnection rate, since the
measurement of the current sheet's length can only be established very
roughly. Likewise, we cannot be sure we have truly resolved the current sheet
itself in our observations, thus our measurement of its width could
well be an overestimate.

On the other hand, there is little evidence that the current sheet is
significantly shorter than the length we report here --- certainly not
orders of magnitude shorter --- and the analysis presented in
\citet{2015SSRv..194..237L} suggests our width measurement is not off by
orders of magnitude either. Further, any error in these two
measurements would have an offsetting effect since both the length and
the widths we report here are upper limits. All of this suggests this
estimate is a reasonable --- if not extraordinarily precise --- one.

\section{Discussion} \label{sec:disc}

Here we have reported on the formation of a long, thin, hot structure
in the wake of a coronal eruption. We believe this is an example of a
direct observation of the kind of current sheet that has long been
predicted to be an essential feature of many solar eruptions.

One question we investigated was whether there was any evidence
  of a thermal halo surrounding the sheet, which has been predicted by several authors \citep{1997ApJ...474L..61Y,
2001ApJ...549.1160Y, 2009ApJ...701..348S, 2016ApJ...823..150T} to form
as a result of thermal conduction out of the current sheet and into
the surrounding plasma. In simulations that include strong thermal conduction, the conduction
substantially increases the density in the current sheet because it
cools the plasma substantially. What this means observationally is
that --- at least in some cases --- a current sheet might be
considerably easier to detect than its surrounding halo of heated
plasma.

In this case, we find little evidence that thermal conduction
transports a significant amount of energy out of the current sheet
early in the event, even as it acts to distribute heat along the sheet
to create the highly uniform, hot layer we discussed in
Section~\ref{subsec:DEM}. 

One possibility is that conduction is simply not as important an
  effect as some authors have predicted. Indeed, this is a real possibility: In a study of the structure
  of a slow magnetosonic shock using two-fluid hydrodynamic equations
  \citet{2010ApJ...718.1491L} found that in some instances, in a low-beta plasma, the
  conduction front might not extend into a broad region beyond the
  current sheet, but could be contained entirely within the reconnection
  outflow jet. In principle, such an effect would reduce the importance of thermal
  conduction and strongly limit the breadth of any thermal halo.

However, later in the event we studied, a broad, diffuse structure
that could be heated by conduction grows at the bottom end of the current sheet. Since we see strong
evidence of shrinking loops in this region, we conclude that this is
probably not part of the narrow Sweet-Parker reconnection region of
the current sheet itself, but rather represents the region of outflow
from the sheet. Thus this is not necessarily incompatible with
  \citeauthor{2010ApJ...718.1491L}'s results.

On the other hand, in a study of DEMs of the cusp-shaped flare regions in several events
including this one, \citet{2015SoPh..290.2211G} reported
temperatures between 11 and 19~MK in the cusp region. They argue that
the high temperatures measured in flare cusps supports the interpretation that
slow shocks at the tips of current sheet outflow play a significant
role in heating flare plasma.

If this latter interpretation is correct, it might help explain why the
current sheet remains so narrow for so long, and, further, might explain its comparatively cooler
temperature. This would be consistent with the analysis of
\citet{1997ApJ...474L..61Y}, who found that thermal conduction plays
an especially important role in the slow shocks region, where
conduction allows heat to flow across the shock and into the inflow
region. Since heat conduction across field lines is comparatively
poor, and the field lines in the center of the sheet are generally
oriented almost parallel to the sheet itself, conduction's central role in
the current sheet region would be to spread heat evenly throughout the
sheet, rather than into the inflow region that surrounds the
sheet. Meanwhile, the slow shocks discussed by
\citeauthor{2015SoPh..290.2211G} border the cusp region where
Petsheck-like reconnection outflow occurs, heating the flare-associated plasma
low in the corona more efficiently, and driving the formation of the
broad cusp-shaped structure itself.

This, coupled with the comparatively low reconnection rate we
measured, indeed suggests the presence of a long and narrow Sweet-Parker
reconnection region bounded by slow shocks at its tips. In their
analysis of the tether-cutting reconnection at the onset of this
event, \citeauthor{2014ApJ...797L..15C} did not report a reconnection
rate, but we can surmise it must have been very fast to facilitate the rapid
escape of the flux rope and the extremely impulsive heating that caused the
X-ray brightness to jump from background levels to the X4.9 flare peak
in just minutes. The rate we measure, about 45 minutes after the onset
of the event, is much slower.

One might speculate that what we are observing in the growth of the
current sheet during this event is a clue to the self-limiting nature of
reconnection during many eruptive flares, in contrast to the rare
events in which reconnection persists for as long as days
after the onset of the eruption \citep[see, for example][and
references therein]{2015ApJ...801L...6W}. Even if the reconnection
rate at the event's onset is very fast, if the eruption escapes quickly enough,
reconnection in its wake cannot keep pace with the rate growth of the
current sheet or the rate of inflow into the sheet, and
the current sheet will lengthen. The lengthening of the sheet creates
a bottleneck, with an increasing amount of flux being advected into
the sheet along its length, but an insufficient increase in the width
of the outflow region that would help to facilitate the flow of
additional field through the current sheet. In this scenario the
reconnection rate will necessarily decrease, eventually becoming so
slow as to effectively switch off the process altogether.

Unfortunately, since high-quality observations of the upper tip of the current
sheet were only possible at a single time, profiling the growth of
this feature --- and thus characterizing the reconnection rate --- over its lifetime was not possible. This is
related partially to an unfortunately timed data gap in the LASCO~C2
coronagraph data, but it is also related to AIA's one major
limitation: its narrow field of view. Further studies of the dynamics
of eruptive flare reconnection might be significantly enhanced with
observations from EUV imagers with larger fields of view. One such
imager, the Sun Watcher with Active Pixels and Image Processing
\citep[SWAP;][]{2013SoPh..286...43S, 2013SoPh..286...67H} onboard ESA's
PROBA2 spacecraft has already demonstrated the value of larger fields of view
in EUV imagers \citep[see, for example,][]{2013ApJ...777...72S, 2014SoPh..289.4545B,
2015ApJ...801L...6W, 10.3389/fspas.2016.00014}. Unfortunately, SWAP,
which has a single 17.4-nm passband with weak response to
high-temperature flare emission, is not ideally tuned for the study of
flares like the one we consider in this paper. 

One imager that might be able to help close the observation gap,
however, is the Solar Ultraviolet Imager
\citep[SUVI;][]{Galarce:EUVModeling:2010, 2013OptEn..52i5102M} to fly later in 2016 on the joint
NOAA/NASA GOES-R series of spacecraft. SUVI will bring together six
EUV passbands --- 93, 131, 171, 195, 284, and 304~\AA\ --- with an
extended field of view similar to SWAP's, and may help make a
significant improvement in our ability to study large structures
associated with eruptions that are presently unobservable. With four
separate (identical) instruments slated to fly on four separate
platforms, SUVI's operational lifetime will likely span two decades,
covering almost two complete solar cycles, so it may play an important
role in answering questions about the nature of reconnection in solar
flares.

There are other tantalizing hints of phenomena --- SADs, shrinking
loops, and inflow-outflow pairs, for example --- that have long been
associated with magnetic reconnection events, but remain, overall,
only partially understood. On the one hand, the fact that we have not
managed to process the data to allow a more quantitative assessment of
the nature of these phenomena is discouraging. On the other hand,
their presence in an apparently clear detection of a current sheet is
strong evidence that our present understanding, that all of these are
manifestations of one fundamental process, is probably correct. Future
observations, perhaps with an imager such as SUVI or other large
field-of-view EUV imagers that have been proposed \citep[][for
example]{2016SPD....47.0805G} may one day help in unraveling the many
remaining questions about the nature of magnetic reconnection in eruptive flares. 

\acknowledgments

We thank Kathy~Reeves for assistance with the DEM analysis tools we
used in the preparation of this paper. We thank the anonymous
  referee for a thoughtful reading of our paper and a number of
  constructive remarks that helped us improve our
  original draft. We also thank Matthew~West and
Elke~D'Huys for thoughtful and productive discussions at the outset of
this project. AB contributed to this research as a
participant in The University of Colorado's NSF-sponsored Research
Experiences for Undergraduates (REU) program, funded by NSF Award
1157020, during the summer of 2016.

\vspace{5mm}
\facility{SDO, SOHO}





\bibliographystyle{../apj} 
\expandafter\ifx\csname natexlab\endcsname\relax\def\natexlab#1{#1}\fi

\end{document}